\documentclass[twoside]{hs04-preprint}
\def\runauthor{S. Vok\'al, S. Lehock\'a, G. I. Orlova, A. Krav\v c\'akov\'a, A. Vok\'alov\'a}
\def\shorttitle{Pecularities in Produced Particles Emission}
\begin{document}

\title{PECULARITIES IN PRODUCED PARTICLES EMISSION \\
IN $^{208}$P\lowercase{b}+A\lowercase{g}(B\lowercase{r}) 
INTERACTIONS AT 158~A~G\lowercase{e}V/\lowercase{c}}

\author{S. Vok\'al, A. Vok\'alov\'a}
\address{VBLHE JINR Dubna, Russia\\
         E-mail: vokal@sunhe.jinr.ru}

\author{A. Krav\v c\'akov\'a, S. Lehock\'a}
\address{University of P.~J.~\v{S}af\'arik, Ko\v{s}ice, Slovakia}

\author{G. I. Orlova}
\address{Lebedev Institute of Physics, RAS, Moscow, Russia}

\maketitle

\abstracts{The angular structures of particles produced in ${}^{208}$Pb induced collisions with Ag(Br) nuclei in an emulsion detector at 158~A~GeV/c have been investigated. Nonstatistical ring-like substructures in azimuthal plane of the collision have been found and their parameters have been determined. The indication on the formation of the ring-like substructures from two symmetrical emission cones -- one in the forward and other in the backward direction in the center-of mass system have been obtained. The ring-like substructures parameters have been determined.  The experimental results are in an agreement with I.~M.~Dremin idea, that mechanism of the ring-like substructures formation in nuclear collisions is similar to that of Cherenkov electromagnetic radiation.
}

\section{Introduction} 

During last two decades different nuclei have been accelerated to relativistic energies and brought to collisions with a great variety of target nuclei. An important aim of investigating nucleus – nucleus collisions is to search for a phenomena connecting with large densities obtained in such nuclear collisions. 

One of the proposed signals for collective phenomena in relativistic nuclear interactions is large fluctuation in the number of produced particles in local regions of phase space \cite{rr01}. As an example, the transition from the QGP (quark-gluon plasma) back to the normal hadronic phase is predicted to give rise to such fluctuations \cite{rr02}.

Whenever it is analyzed, the observed effects are dominated by statistical fluctuations. Significant deviations from them are only observed after painstaking efforts to remove the statistical part of the fluctuations \cite{rr03}. Several such attempts have been made in past.

In analyses of an azimuthal distributions of produced particles two different classes of substructures were revealed, which could be referred to as jet-like and ring-like structures. The ring-like structures are occurrences where many particles are produced in a narrow region along the rapidity axis, which at the same time are diluted over the whole azimuth. The jet-like structures consist of cases where particles are focused in both dimensions \cite{rr04}. 

For the first time the individual nucleus-nucleus collisions with a ring-like substructure of produced particles in the azimuthal plane have been observed more then 20 years ago in cosmic ray experiments \cite{rr05}. Later a lot of the nucleus-nucleus collisions with the ring-like substructure were observed in the accelerator experiments at high energy \cite{rr03,rr06,rr07,rr08}.

A new mechanism of multiparticle production at high energies was proposed in \cite{rr09,rr10,rr11}. This mechanism is similar to that of Cherenkov electromagnetic radiation. As a hadronic analogue, one may treat an impinging nucleus as a bunch of confined quarks each of which can emit gluons when traversing a target nucleus \cite{rr12,rr13}. The idea about possible Cherenkov gluons is relying \cite{rr09} on experimental observation of the positive real part of the elastic forward scattering amplitude of all hadronic processes at high energies. This is a necessary condition for such process because in the commonly used formula for the refractivity index its excess over 1 is proportional to this real part. Later I.M.Dremin \cite{rr10} noticed that for such thin targets as nuclei the similar effect can appear due to small confinement length thus giving us a new tool for its estimate. If the number of emitted gluons is large enough and each of them generates a mini-jet, the ring-like substructure will be observed in the target (azimuthal) diagram. If the number of emitted gluons is not large, we will see several jets correlated in their polar, but not in the azimuthal angles. Central collisions of nuclei are preferred for observation of such effects because of a large number of participating partons.

In the present study the ring-like substructures of charged produced particles from $^{208}$Pb + Ag(Br) interactions in emulsion detector at 158 A GeV/c have been analyzed.

\section{Experiment}

The stacks of NIKFI BR-2 nuclear photoemulsions have been irradiated horizontally by $^{208}$Pb beam at 158 A GeV/c (the CERN SPS accelerator - experiment EMU12) \cite{rr14}. The photoemulsion method allows to measure: 

\begin{itemize}
\item[-]
  multiplicities of any charged particles: produced particles $(N_s)$ with $\beta > 0.7$, 
  projectile fragments $(N_F)$ with $\beta \approx 0.99$ and target fragments $(N_h)$ 
  with $\beta < 0.7$;
\item[-]
  angles of particles with the resolution of $\Delta \eta = 0.010 - 0.015$ rapidity units in the 
  central region, pseudo-rapidity is given by $\eta = - ln(tan(\Theta/2))$, and $\Theta$ is the 
  emission angle with respect to the beam direction,  
\item[-]
  charges of projectile fragments $Z_F$. 
\end{itemize}

In this work we have analyzed 628 $^{208}$Pb+Em collisions found by the along-the track scanning. From the collisions we have selected three centrality groups determined by the multiplicity of the shower particles: $350 \le N_s < 700, 700 \le N_s < 1000$ and $N_s \ge 1000$.

\begin{figure}[!thb]
\epsfysize=8cm
\epsfxsize=8cm
\begin{center}
\epsffile{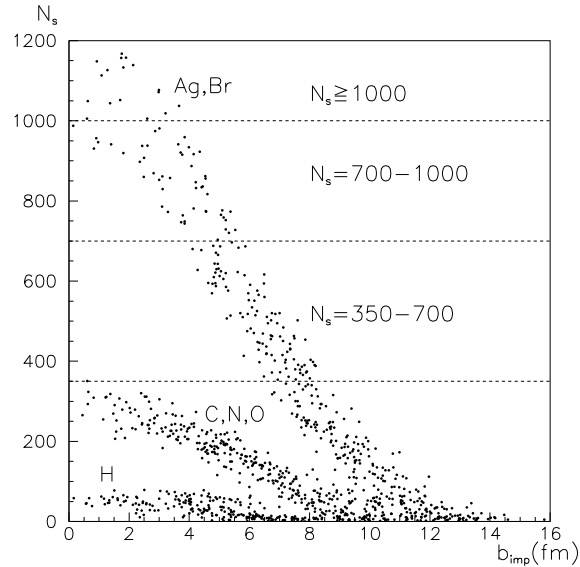}
\end{center}
\caption[*]{ $N_s$ dependence on $b_{imp}$  for Pb+Em collisions at 158 A GeV/c for different mass of photoemulsion components. }
\label{ff01}
\end{figure}

The dependence of the number of produced particles $N_s$ on the impact parameter $b_{imp}$ calculated by the FRITIOF model \cite{rr15} for our experiment is shown in the Fig. \ref{ff01} for different mass of photoemulsion components $A_t$.

One can see that the criterion $N_s \ge 350$ selects the interactions of lead nuclei with the heavy emulsion targets Ag and Br with $b_{imp} < 8 fm$ only. Moreover the group with $N_s \ge 1000$ comprises the central Pb+Ag(Br) interactions with impact parameter $b_{imp} \approx (0-2) fm$. 
The mean values of the shower particles multiplicity and pseudorapidity in the sample are $\langle N_s \rangle_{exp} = 623 \pm 30$ $(\langle N_s \rangle_{fritiof} = 649)$ and $\langle \eta \rangle _{exp}= 3.7 \pm 0.2$ ($\langle \eta \rangle _{fritiof}= 3.6$).

\section{Method}

A method we use to search for a ring-like substructure and to determine parameters they have was devised in our paper \cite{rr03}. The multiplicity $N_d$ of analyzed subgroup from an individual event is kept a fixed. Each $N_d$ - tuple of consecutive particles along the $\eta$ - axis of individual event can then be considered as a subgroup characterized by:
\begin{itemize}
\item[a)] a size
\begin{equation} \label{eq01}
\Delta\eta_d = \eta_{max} - \eta_{min}
\end{equation}
where $\eta_{min}$ and $\eta_{max}$ are the pseudorapidity values of the first and last particles in the subgroup;
\item[b)] a density 
\begin{equation} \label{eq02}
\rho_d = \frac{N_d}{\Delta\eta_d}
\end{equation}
\item[c)] an average pseudorapidity (or a subgroup position) 
\end{itemize}
\begin{equation} \label{eq03}
\eta_m = \frac{\sum\eta_i}{N_d}.
\end{equation}

To parameterize the azimuthal structure of the subgroup in a suitable way a parameter of the
azimuthal structure 
\begin{equation} \label{eq04}
S_2 = \sum{(\Delta\Phi_i)}^2
\end{equation}
have been suggested, where $\Delta\Phi$ is the difference between azimuthal angels of two neighboring particles in the investigated group (starting from the first and second and ending from the last and first). For the sake of simplicity it was counted $\Delta\Phi$ in units of full revolutions $\sum(\Delta\Phi_i) = 1$.
 
The parameter $S_2$ is large $(S_2 \to 1)$ for the particle groups with the jet-like structure 
and is small $(S_2 \to 1/N_d)$ for the particle groups with the ring-like structure. The expectation value for the parameter $S_2$, in the case of stochastic scenario with independent particles in the investigated group, can be analytically expressed as
\begin{equation} \label{eq05}
\langle S_2 \rangle = \frac{2}{N_d + 1}.
\end{equation}
This expectation value can be derived from the distribution of gaps between neighbors.

\begin{figure}[!bht]
\epsfysize=8cm
\epsfxsize=8cm
\begin{center}
\epsffile{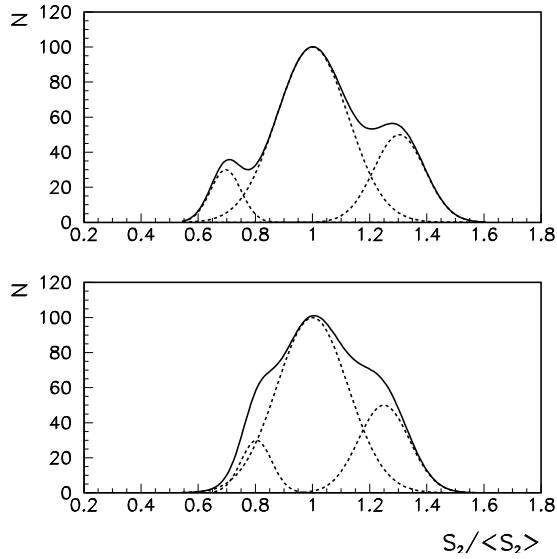}
\end{center}
\caption[*]{ The examples of summary $S_2/ \langle S_2 \rangle$ - distribution from three effects: stochastic distribution  +  ring-like effect distribution   +  jet-like effect distribution. }
\label{ff02}
\end{figure}

What can one wait to see in the experimental $S_2/ \langle S_2 \rangle$ – distributions in different scenarios? It is illustrated schematically in Fig.~\ref{ff02} using for example Gauss distributions. In case of a pure stochastic scenario the distribution would have a peak position at 1. The existence of the jet-like substructures in collisions results to appearance of additional $S_2$ distribution from this effect, but shifted to the right side in comparison with stochastic distribution. Analogously, the existence of the ring-like substructures results to appearance of additions to $S_2$ distribution from this effect but shifted to the left side. As result, the summary $S_2$ distribution from this three effects may have different form depends of mutual order and sizes.

\begin{figure}[!b]
\epsfysize=7cm
\epsfxsize=7cm
\begin{center}
\epsffile{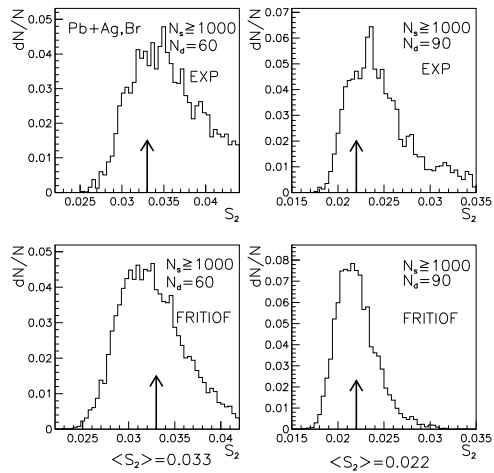}
\end{center}
\caption[*]{ The experimental (up) and calculated by the FRITIOF (down)
 $S_2$ – distributions for  $^{208}Pb+Ag(Br)$ collisions with $N_s \ge 1000$ at 158 A GeV/c
 and two different multiplicities in the subgroup $N_d = 60$ and $N_d = 90$.}
\label{ff03}
\end{figure}

\section{Results}

The first detailed study of the average values of the parameter $S_2$ was performed in \cite{rr03}. The azimuthal substructures of particles produced within dense and dilute groups along the rapidity axis in the central $^{16}O$ and $^{32}S$ induced collisions with Ag(Br) and Au targets at 200 A GeV/c (EMU01 data sets) were analyzed. The results were compared with different FRITIOF calculations including $\gamma$ – conversion and HBT effects. It was conclude that jet-like and ring-like events do not exhibit significant deviations from what can be expected from stochastic emission. But average values study don't give a full information about an effect.

The detail study of ring-like substructures of the particles produced in 11.6 A GeV/c $^{197}Au$ interactions with Ag(Br) targets in emulsion detector has been done in \cite{rr16}.  Nonstatistical ring-like substructures have been found and cone emission angles as well as other parameters have been determined.

Fig. \ref{ff03} shows the $S_2$ - distributions for $^{208}Pb+Ag(Br)$ collisions at 158 A GeV/c with $N_s \ge 1000$ and two different subgroups with $N_d = 60$ and $N_d = 90$. The experimental $S_2$ - distributions are on the top of the figure and the corresponding ones given by the FRITIOF model are down.

\begin{figure}[!b]
\epsfysize=7.3cm
\epsfxsize=7.3cm
\begin{center}
\epsffile{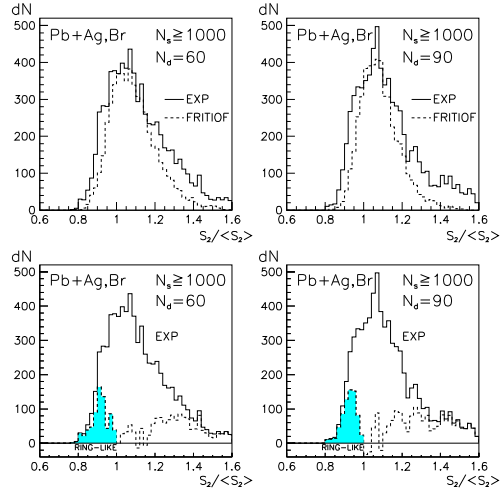}
\end{center}
\caption[*]{ The $S_2/ \langle S_2 \rangle$ distributions – experimental and FRITIOF (up) and the experimental $S_2/ \langle S_2 \rangle$ distributions after the subtraction of the FRITIOF’s ones (down) for $N_d = 60$ (left) and $N_d = 90$ (right). }
\label{ff04}
\end{figure}

\begin{figure}[!t]
\epsfysize=7.5cm
\epsfxsize=7.5cm
\begin{center}
\epsffile{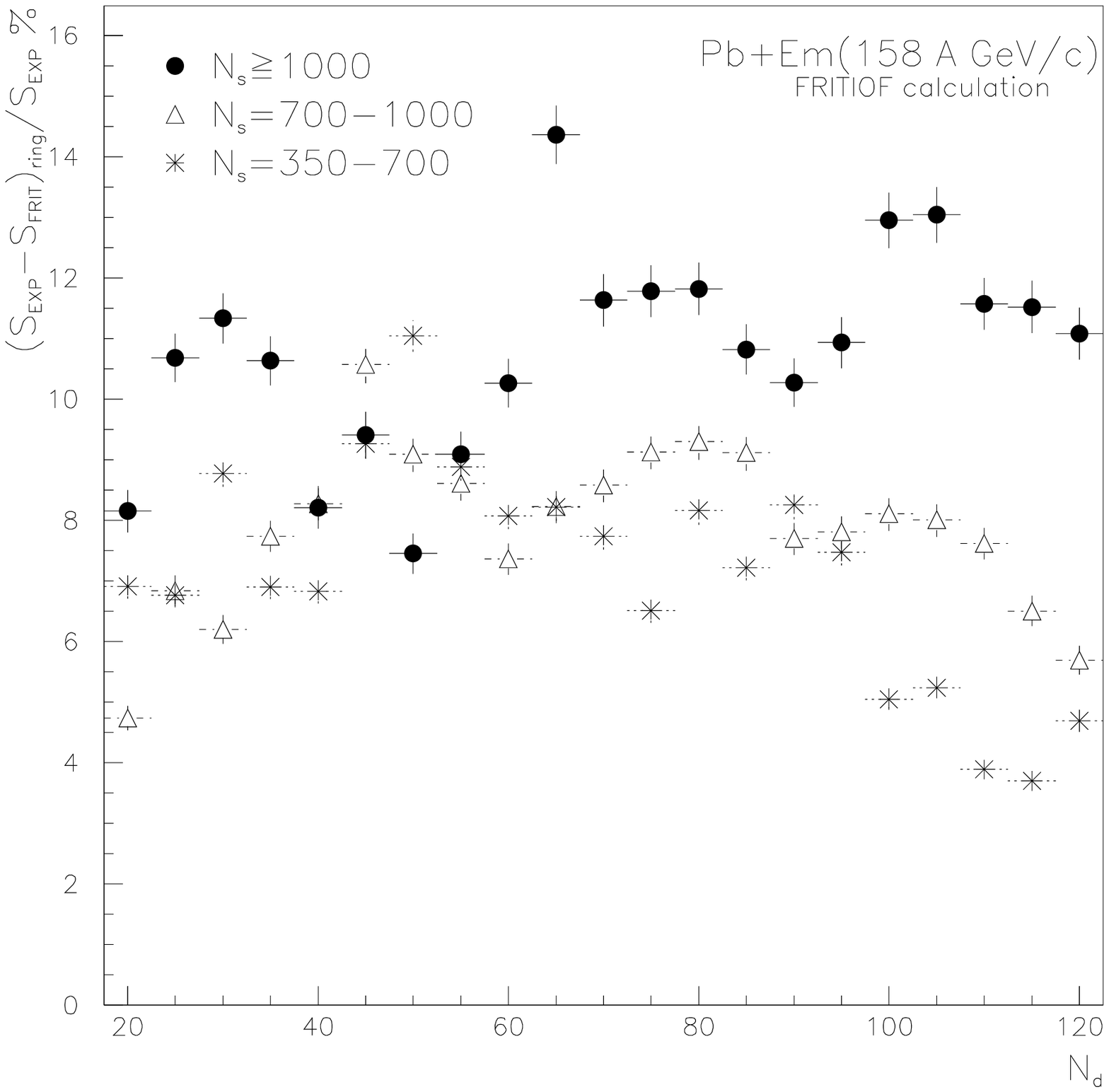}
\end{center}
\caption[*]{ The dependence of the contribution of the ring-like effect (\%) to the experimental data as a function of the number $N_d$ of the particles in a subgroup for three groups of $N_s$. }
\label{ff05}
\end{figure}

\begin{figure}[]
\epsfysize=8cm
\epsfxsize=8cm
\begin{center}
\epsffile{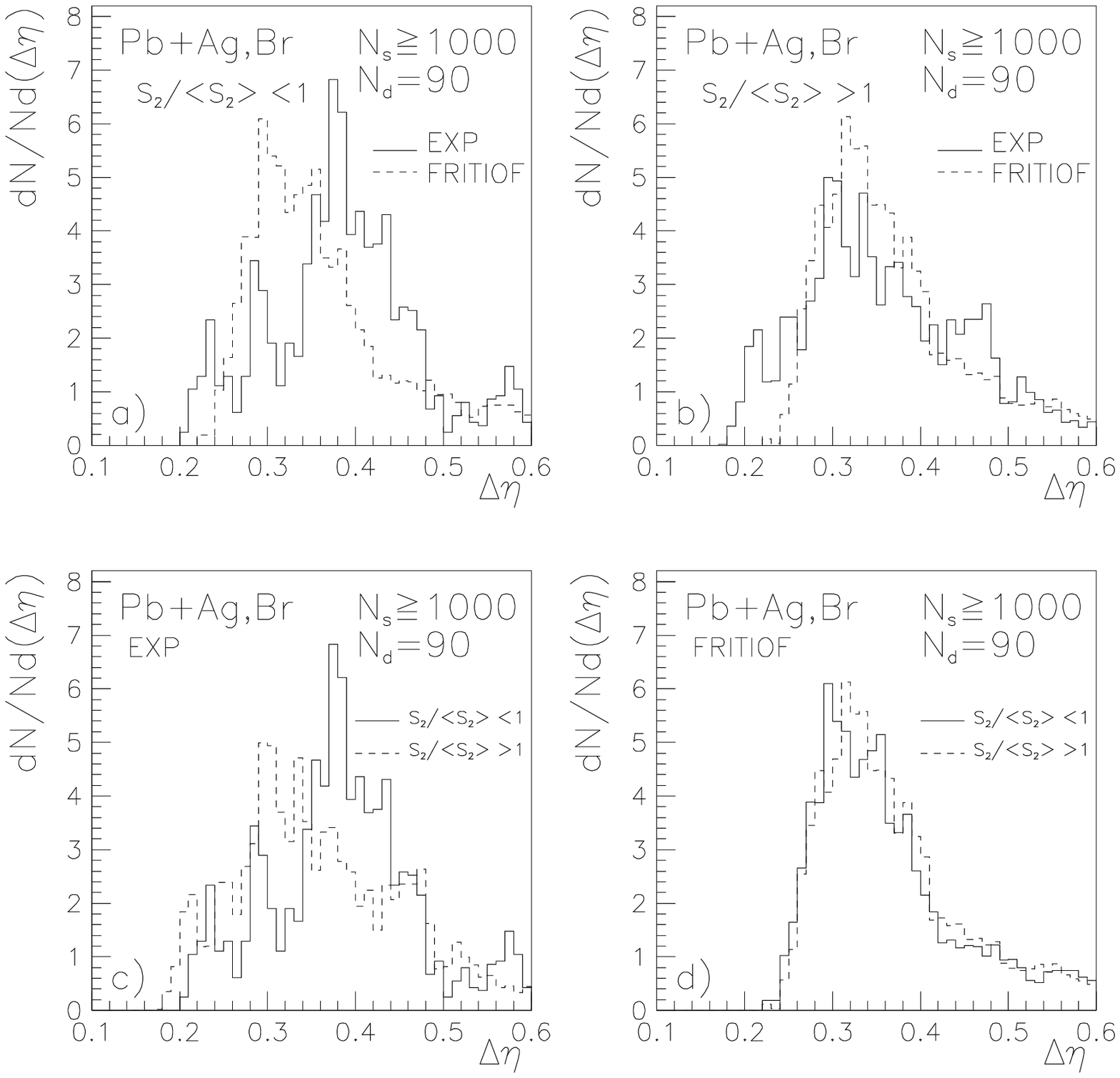}
\end{center}
\caption[*]{ Comparison the experimental and FRITIOF model $\Delta\eta$ - distributions for collisions with $N_s \ge 1000$ and subgroups with $N_d = 90$ for two regions: $S_2/ \langle S_2 \rangle < 1$ and $S_2/ \langle S_2 \rangle > 1$. }
\label{ff06}
\end{figure}

In case of a pure stochastic scenario $S_2$ - distribution would have a peak position around $S_2 = 0.033$ for $N_d = 60$ and around $S_2 = 0.022$ for $N_d = 90$ (see equation (\ref{eq05})). As it was illustrated in Fig. \ref{ff02} the existence of the jet-like substructures in collisions results to lengthening of the right part of the $S_2$ - distribution and to appearance of additional peaks there. The existence of the ring-like substructures results to deformations and/or to appearance of additional peaks at the left part of the $S_2$ - distribution, where $S_2 < \langle S_2 \rangle$. 

One can see that the experimental distributions are shifted to the right, have a tail in the right part and are broader than the spectra calculated by the FRITIOF. The left parts of both experimental distributions are not as smooth as in the model and there are some shoulders that refer to the surplus of the events in this region.

The experimental normalized $S_2/ \langle S_2 \rangle$ - distributions compared with the distributions calculated by the FRITIOF model are presented on the top of the Fig. \ref{ff04}. The model distributions were aligned according to the position of the peak with the experimental one. The FRITIOF model includes neither the ring-like nor the jet-like effects, so the model distributions are used like the statistical background.

Below in the Fig.\ref{ff04} there are the results obtained from the experimental data after the subtraction of the statistical background. The resultant distributions have two very good distinguishable hills, the first in the region $S_2/ \langle S_2 \rangle < 1$, where the ring-like effects are expected and the second in the jet-like region -  $S_2/ \langle S_2 \rangle > 1$. The probability of the formation of the ring-like substructures can be estimated as a rate of the surface of the ring-like part to the full surface of the experimental distribution.

The estimation of the contribution of the ring-like effect to the experimental data as a function of the number $N_d$ for three groups of $N_s$ is presented in Fig. \ref{ff05}. These dependences for every $N_s$ group have a several maxima situated nearly at the same value of $N_d$. In the group with $N_s \ge 1000$ this dependence slowly increases as a full, in the two other groups the dependences are decreasing.

 The shape of these distributions can be explained as follows. The observed ring-like substructures seem to be a sum result of independent formation of the rings from several individual gluons, a number of which may be different. Each hill in dependences in Fig. \ref{ff05} corresponds to ring-like substructures with different multiplicity of the number of formatting gluons. The average multiplicity of the gluons $\langle N_g \rangle$, participated in the ring-like process formation is proportional to the overlap of the collided nuclei, i.e. it has to be bigger for more central interaction. So, the average multiplicity $\langle N_d \rangle$ of particles in the ring-like substructure is not fixed, but it is proportional to the number of the gluons participated in the formation of the ring-like substructures.

The next step was to investigate the ring-like subgroups size $\Delta\eta$. The Fig.~\ref{ff06} shows the $\Delta\eta$ distribution for the most central interactions and the number of particles in the subgroup $N_d = 90$. This distribution is calculated for two classes $S_2/ \langle S_2 \rangle < 1$ – the region of the ring-like effects and $S_2/ \langle S_2 \rangle > 1$ – the region of the jet-like effects. The experimental data are compared with calculated the FRITIOF model calculations. One can see that there are some distinctions in the shapes of the experimental and model distributions from one side (Fig.~\ref{ff06}a,b) and in the shapes of the experimental distributions for the regions $S_2/ \langle S_2 \rangle < 1$ and $S_2/ \langle S_2 \rangle > 1$ from the other side (Fig.~\ref{ff06}c). There are no differences in the distributions calculated by the model for both classes (Fig.~\ref{ff06}d). There appeared 3 or 4 peaks in the experimental $\Delta\eta$ – distribution in the ring-like effect region $S_2/ \langle S_2 \rangle < 1$, that we don’t see in other cases. This fact  may means that  this peaks corresponds to sizes of ring-like substructions.

\begin{figure}[!t]
\epsfysize=4.5cm
\epsfxsize=14cm
\begin{center}
\epsffile{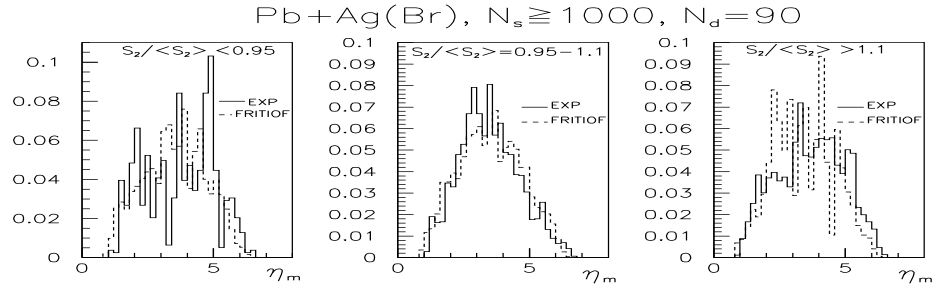}
\end{center}
\caption[*]{ The ring-like subgroup position $\eta_m$ for $N_s \ge 1000$, $N_d = 90$ and different $S_2/ \langle S_2 \rangle$ classes is given for experimental data and FRITIOF model. }
\label{ff07}
\end{figure}

To analyze the ring-like subgroup position $\eta_m$ on the pseudorapidity axis, the $\eta_m$ - distributions for $N_s \ge 1000$ and $N_d = 90$ are given for experimental data and FRITIOF model in Fig.~\ref{ff07}. For comparison are these distributions divided into three classes:
\begin{itemize}
\item[] $S_2/ \langle S_2 \rangle < 0.95$ - the region of the ring-like effects,
\item[] $S_2/ \langle S_2 \rangle = 0.95-1.10$ - statistical background,
\item[] $S_2/ \langle S_2 \rangle > 1.10$ - the region of the jet-like effects.
\end{itemize}

The difference with the FRITIOF model calculations in the $\eta_m$ - distributions in ring-like region $S_2/ \langle S_2 \rangle < 0.95$ indicates to existence of two $\eta_m$ - regions of preferred emission of ring-like substructures $-$ one in the forward and second in the back directions in center-of-mass system.

If the ring-like substructures have been appeared due to an effect analogous to Cherenkov light there may be in a collision two ring-like subgroups forming two cones with the equal emission angles – one in the forward and another in the backward direction in center-of-mass system. As well known the cone emission angle of the Cherenkov light is directly connected with the refractive index of matter, in our case of nuclear matter, and it is a way to measure it. It is interesting to note that the refractive index of nuclear matter may be changed in the case of the changes the nuclear matter properties, for example, in the case of phase transition from a normal hadronic matter to quark qluon-plasma.

\section*{Conclusion}

The azimuthal ring-like substructures of produced particles from collisions induced by the 158 A GeV/c $^{208}Pb$ beam in the emulsion detector have been investigated.
\begin{itemize}
\item The additional interactions in the region of the ring-like substructures ($S_2/ \langle S_2 \rangle < 1$) in comparison to the FRITIOF model calculations have been observed.
\item The $\Delta\eta$ – distribution, which gives the information about a subgroup size in pseudorapidity scale, for the experimental data in ring-like region ($S_2/ \langle S_2 \rangle~<~1$) differs from the others.
\item The difference with the FRITIOF model calculations in the $\eta_m$ – distributions in ring-like region ($S_2/ \langle S_2 \rangle < 0.95$) indicates to existence of two $\eta_m$ - regions of preferred emission of ring-like substructures $-$ one in the forward and second in the back directions in center-of-mass system.
\end{itemize}
Our results are in good agreement with mechanism of the ring-like substructures production similar to that of Cherenkov electromagnetic radiation.

\section*{Acknowledgements}This work was supported by Science and Technology Assistance Agency under the contract No. APVT-51-010002 and by RFBR grants No. 03-02-17079 and No. 04-02-170079.

\end{document}